\documentclass[11pt]{article}
\usepackage{epsfig,apjfonts}
\def\gtrsim{\mathrel{\hbox{\rlap{\hbox{\lower4pt\hbox{$\sim$}}}\hbox{$>$}}}}
\def\lesssim{\mathrel{\hbox{\rlap{\hbox{\lower4pt\hbox{$\sim$}}}\hbox{$<$}}}}
\begin{document}

\vspace*{-0.50in}
\begin{center}
{\Large\bf 
The Coronal Physics Investigator (CPI) Experiment for ISS:

\vspace*{0.03in}
A New Vision for Understanding Solar Wind Acceleration}

\vspace*{0.06in}
J.~L.~Kohl,$^{1}$
S.~R.~Cranmer,$^{1}$
J.~C.~Raymond,$^{1}$
T.~J.~Norton,$^{1}$
P.~J.~Cucchiaro,$^{2}$
D.~B.~Reisenfeld,$^{3}$
P.~H.~Janzen,$^{3}$
B.~D.~G.~Chandran,$^{4}$
T.~G.~Forbes,$^{4}$
P.~A.~Isenberg,$^{4}$
A.~V.~Panasyuk,$^{1}$
and
A.~A.~van~Ballegooijen$^{1}$

\vspace*{0.02in}
{\em
$^{1}$Harvard-Smithsonian CfA,
$\,$
$^{2}$L-3 Com IOS,
$\,$
$^{3}$U.~Montana,
$\,$
$^{4}$U.\  New Hampshire
$\,$
}

\vspace*{0.14in}
{\bf Abstract}
\end{center}

\vspace*{-0.05in}
In February 2011 we proposed a NASA Explorer Mission of Opportunity
program to develop and operate a large-aperture ultraviolet coronagraph
spectrometer called the Coronal Physics Investigator (CPI) as an
attached International Space Station (ISS) payload.
The primary goal of this program is to identify and characterize
the physical processes that heat and accelerate the primary and
secondary components of the fast and slow solar wind.
In addition, CPI can make key measurements needed to understand CMEs.
CPI is dedicated to high spectral resolution measurements of the
off-limb extended corona with far better stray light suppression
than can be achieved by a conventional instrument.

UVCS/{\em{SOHO}} allowed us to identify what additional measurements need
to be made to answer the fundamental questions about how solar wind
streams are produced, and CPI's next-generation capabilities
were designed specifically to make those measurements.
Compared to previous instruments, CPI provides unprecedented
sensitivity, a wavelength range extending from 25.7 to 126 nm, higher
temporal resolution, and the capability to measure line profiles of
He II, N V, Ne VII, Ne VIII, Si VIII, S IX, Ar VIII, Ca IX,
and Fe X, never before seen in coronal holes above 1.3 solar radii.
CPI will constrain the properties and effects of coronal MHD waves by
(1) observing many ions over a large range of charge and mass,
(2) providing simultaneous measurements of proton and electron
temperatures to probe turbulent dissipation mechanisms, and
(3) measuring amplitudes of low-frequency compressive fluctuations.

CPI is an internally occulted ultraviolet coronagraph that provides the
required high sensitivity without the need for a deployable boom,
and with all technically mature hardware including an ICCD detector.
A highly experienced Explorer and ISS contractor, L-3 Com Integrated
Optical Systems and Com Systems East, will provide the tracking and
pointing system as well as the instrument, and the integration to the ISS.

\vspace*{0.03in}
\begin{center}
{\bf 1. Background and Motivation}
\end{center}

This white paper summarizes the proposed
Coronal Physics Investigator (CPI) experiment.
It is being circulated together with a poster presentation
to be given by Raymond et al.\   at the June 12--16, 2011
meeting of the Solar Physics
Division of the AAS in Las Cruces, New Mexico.

CPI (pictured in Figure 1)
follows on from the discoveries of UVCS/{\em{SOHO}} to answer the
basic question:~{\em{What physical processes heat and accelerate
both the major (proton, electron, helium) and minor (heavy ion)
plasma components of the fast and slow solar wind?}}
In the remainder of this section, we outline the historical context
of CPI and how this experiment will fundamentally transform our
observational understanding of solar wind acceleration.
Section 2 gives additional information about the scientific goals of CPI.
Section 3 describes the CPI science payload, and
Section 4 discusses the accommodation of CPI on the ISS.
Section 5 gives an overview of the CPI mission, and
Section 6 describes the CPI team, management, and costs.
\begin{figure}[!t]
\vspace*{-0.02in}
\hspace*{0.67in}
\epsfig{figure=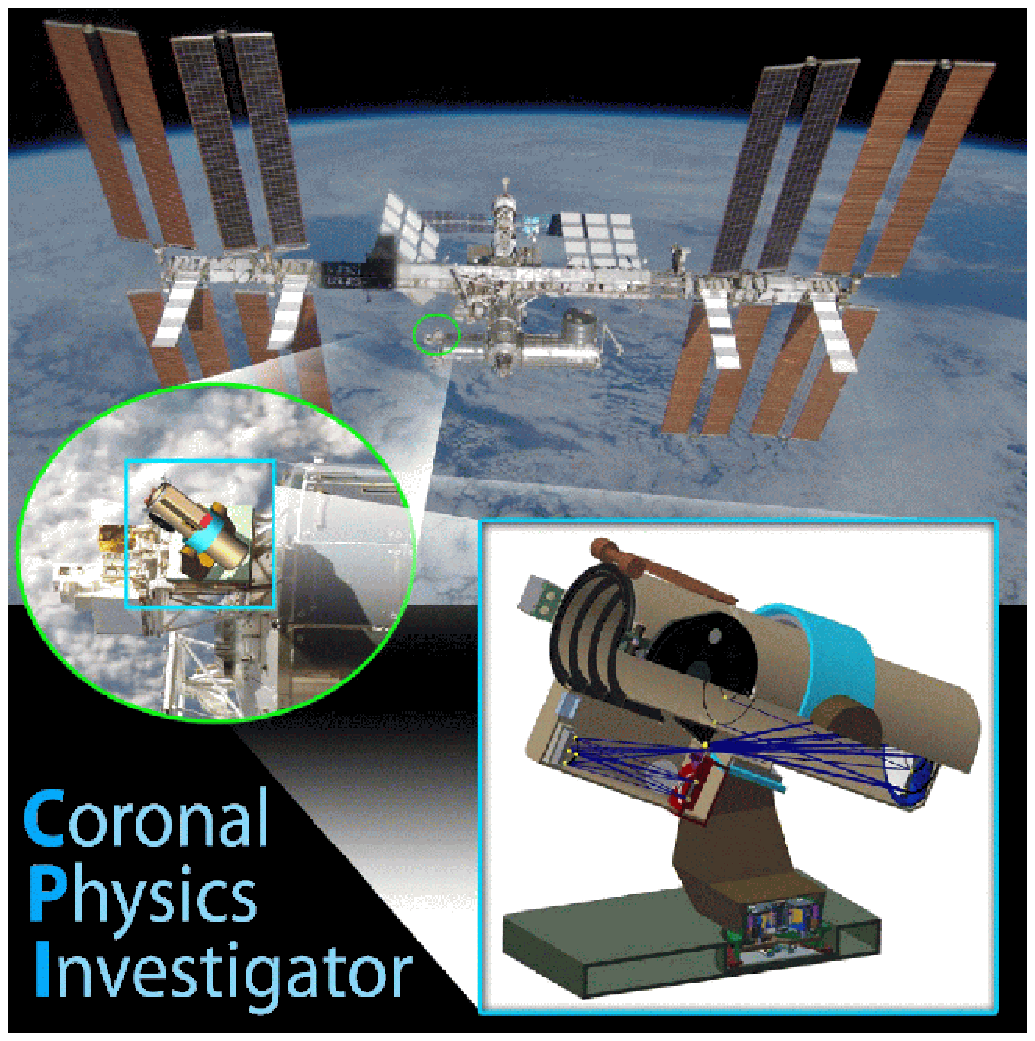,width=4.95in}

\small
{\bf Figure 1:}
CPI has been proposed as a Small Complete Mission for the
International Space Station (ISS).
The baseline location for CPI on ISS is the Columbus
EPF Starboard Overhead Zenith (SOZ) site.
\end{figure}

Ultraviolet spectroscopy of the extended solar corona (defined here
as $\geq 1.4$ solar radii, $R_{\odot}$, from Sun center) has become
a powerful tool for obtaining detailed empirical descriptions of
the regions where solar wind streams and coronal mass ejections (CMEs)
undergo most of their acceleration.
Ultraviolet spectroscopy allows us to determine the velocity 
distributions and outflow velocities of protons and minor ions near
the Sun, it provides absolute chemical abundances and charge state
distributions, and it is
capable of determining the velocity distributions and densities
of electrons (e.g., Withbroe et al.\  1982; Kohl et al.\  2006).
These observations provide the constraints needed to test and guide
theoretical models of extended coronal heating, solar wind acceleration,
and CME production.

Occulted ultraviolet telescope-spectrometers were first
developed and flown on sounding rockets (1979, 1980, 1982) and on
the Shuttle-deployed {\em Spartan 201} payload (1993, 1994, 1995, 1998);
see, e.g., Kohl et al.\  (1978, 1994, 2006).
Following on the successes of these pilot observations, the
Ultraviolet Coronagraph Spectrometer (UVCS)
instrument was flown on the ESA/NASA {\em Solar and Heliospheric
Observatory} ({\em{SOHO}}) spacecraft (Kohl et al.\  1995).
UVCS/{\em{SOHO}} began observing in 1996 and is scheduled to
continue operations until September 2012.

In the fast solar wind, UVCS discovered
remarkably strong preferential ion heating for
O$^{+5}$ and Mg$^{+9}$ and pronounced kinetic anisotropies
(with $T_{\perp} > T_{\parallel}$) in polar coronal holes
(Kohl et al.\  1997, 1998, 2006; Cranmer et al.\  1999, 2008).
These results rekindled theoretical efforts to
understand the heating of the extended corona
by ion cyclotron resonance (Hollweg \& Isenberg 2002),
though we still do not understand
how these waves are produced.
In the slow solar wind, much of which originates in streamers,
efforts have centered on attempts to understand
the relative roles of reconnection and
magnetic flux tube geometry, but there is evidence
for wave processes similar to those seen
in the fast wind (Frazin et al.\  2003).

Despite the advances outlined above, though, the diagnostic
capabilities of UVCS were limited to what was foreseen before
{\em SOHO} (when only H~I Ly$\alpha$ had been observed).
UVCS allowed us to identify what additional measurements need
to be made to answer the fundamental questions about how solar wind
streams are produced, and we have designed the next-generation
capabilities of the CPI experiment specifically to make those
measurements.
As will be described below in more detail,
CPI provides unprecedented sensitivity, a wavelength range
extending from 25.7 to 126 nm, improved time resolution, and the
capability to measure UV line profiles
never before seen in coronal holes in the extended corona. 
CPI would be the first mission to be able to fully characterize the
plasma properties in the region where the solar wind's
primary energy and momentum addition occurs.

The CPI coronagraph spectrometer design is a departure from earlier
next-generation concepts that achieved the required increase in
sensitivity with a remote external occulter mounted on a deployable
boom, and improved detector performance with an
Intensified Active Pixel Sensor (IAPS) detection system with
a high maximum count per pixel capability.
CPI achieves its high sensitivity with an internally occulted
coronagraph design that provides up to a factor of 200 increase in
count rates over UVCS/{\em{SOHO.}}
Since solar disk to solar coronal intensity ratios in the
ultraviolet are smaller than for visible light, it was possible
to design the internally occulted CPI to have sufficient stray light
suppression to observe spectral lines of interest in coronal
holes at the crucial heights of $> 1.8$ $R_{\odot}$, where collisionless
wave-particle interactions are believed to dominate.
CPI uses an ICCD detector similar to those provided for {\em Swift}
and {\em XMM/Newton} by the Mullard Space Science Laboratory.
This detector meets all of the CPI observation requirements, and
eliminates any question about its technical readiness level.
The CPI resource requirements are well within the capabilities of
the ISS, which can provide 16 orbits per day and 310 days per year
of solar observations.
By taking advantage of the existing capabilities of ISS, CPI
achieves its science goals at far less cost than would be needed
for an independent mission.

The CPI project will not only make a tremendous advance in solar
wind science, it will provide a bridge to the next solar maximum
when there is considerable interest in a mission to focus on the
most energetic flares, CMEs, and energetic particle acceleration
events (e.g., the {\em SEE~2020} mission concept; Lin et al.\  2010).
The new CPI team---including Dan Reisenfeld as Project Scientist
and Paul Janzen as Instrument Scientist, as well as Tim Norton
as Project Manager and the new collaboration with the
industrial partner L-3 Com Integrated Optical Systems---would
continue to be in place for this opportunity.

CPI fits well into the constraints of an Explorer Small Complete
Mission with large technical reserves, seven months of costed
schedule reserves beyond a robust 48 month phase B/C/D baseline
development schedule, 25\% cost reserves plus a 5\% margin,
and a \$4M descope plan with vary little impact on the
primary science.

\vspace*{0.03in}
\begin{center}
{\bf 2. Scientific Goals and Objectives}

\vspace*{0.15in}
{\em 2.1. Solar Wind Acceleration}
\end{center}

\vspace*{0.03in}
CPI will address the fundamental question of the identification of
the energy and momentum deposition processes that accelerate
the solar wind.
Figure 2 illustrates the range of 
solar wind issues to be explored by CPI.
A large body of
space-based and ground-based observations suggests that
Alfv\'{e}n waves exist in the lower solar atmosphere and carry enough
power to accelerate the wind and heat the quiet corona
(e.g., Cranmer \& van Ballegooijen 2005;
De Pontieu et al.\  2007; Moore et al.\  2011).
However, the mechanisms that convert these waves into heat and kinetic
energy are still not understood.
The lack of tight constraints on the plasma properties
near the Sun is the major roadblock to a comprehensive understanding
of the relevant processes.
Thus, CPI has been designed to make major breakthroughs by answering
specific questions such as the following.

\begin{figure}[!t]
\vspace*{-0.02in}
\hspace*{0.23in}
\epsfig{figure=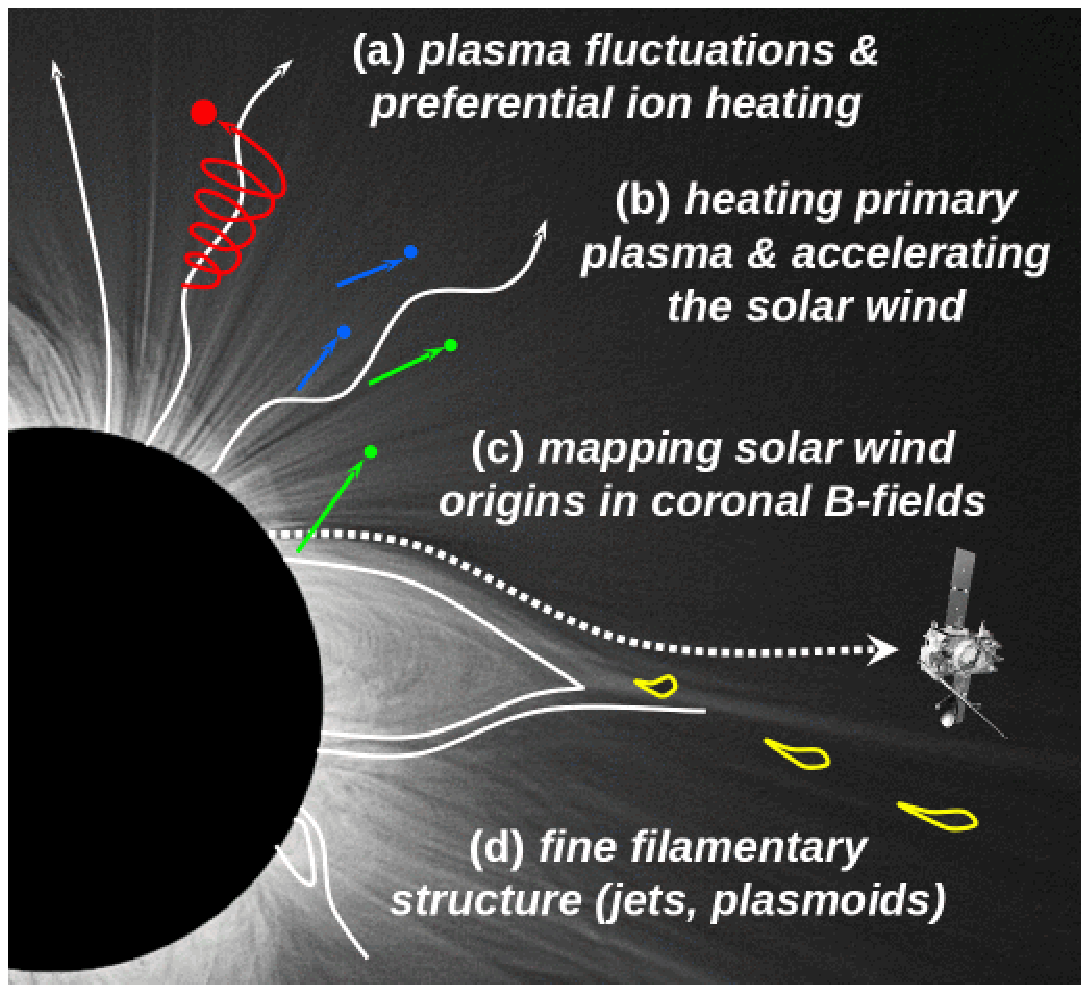,width=2.75in}

\vspace*{-2.55in}
\hspace*{3.16in}
\epsfig{figure=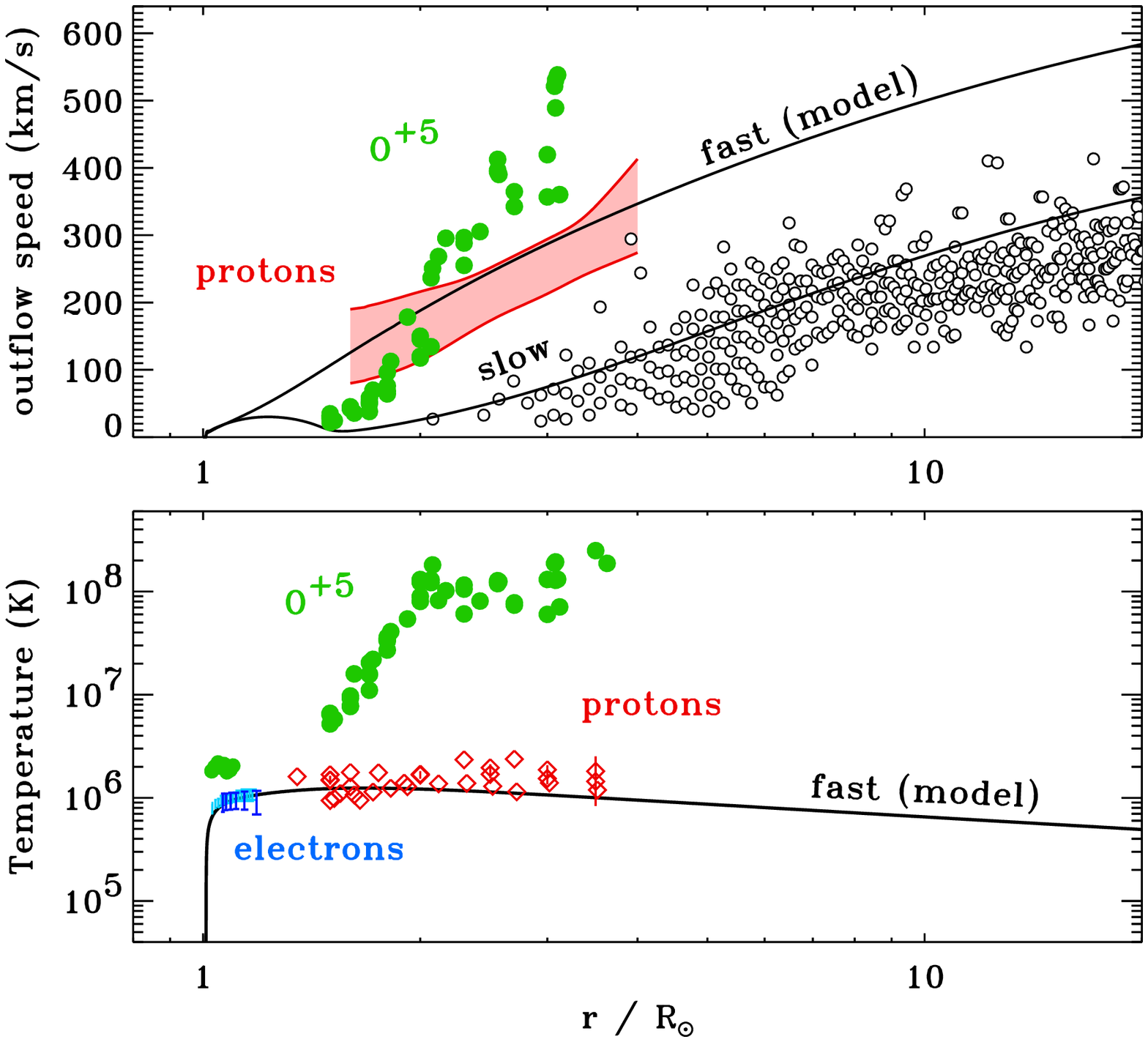,width=2.95in}

\vspace*{-0.01in}
\small
{\bf Figure 2:}
{\em Left:} Eclipse image (Pasachoff et al.\  2007) overlaid with
the solar wind questions addressed by CPI.
{\em Right:} collection of coronal outflow speeds
(top) and temperatures (bottom) from models and
UV spectroscopy and coronagraph measurements.
Black curves are model results from Cranmer et al.\  (2007).
UVCS/{\em{SOHO}} measurements in coronal holes (red: protons,
green: O$^{+5}$) are described in more detail by Cranmer (2009).
Black open circles denote slow wind outflow speeds from
Sheeley et al.\  (1997) and blue symbols denote electron temperatures
from Wilhelm (2006) and Landi (2008).
\end{figure}
{\em How and where do plasma fluctuations
drive the preferential ion heating and acceleration?}
UVCS discovered extreme ion properties in coronal holes
that pointed to the damping of high-frequency (10--$10^4$ Hz)
ion cyclotron waves, but these measurements 
were limited to only two minor ions (Kohl et al.\  1998, 1999).
Thus, it has been difficult to unambiguously confirm the ion cyclotron
idea or distinguish it clearly from other processes
such as stochastic Alfv\'{e}n wave damping (e.g., Voitenko \& Goossens
2004; Chandran 2010), shock acceleration (Lee \& Wu 2000),
or shear-driven instabilities (Markovskii et al.\  2006).
By measuring the properties of ions with a wide range of charge and mass
over a large range of heights, CPI will tightly
constrain the properties of the spectrum of fluctuations
thanks to its high sensitivity and large spectral coverage.
By comparing CPI minor ion measurements to the predictions of kinetic
and MHD models, we can tightly constrain the spectral slope, the
local damping, and the precise modes (Alfv\'{e}n, fast, slow) of
the waves.
As the number of ions observed in each region increases, our confidence
in the uniqueness of a successful theoretical
model (based on a given process) increases as well.

{\em What processes contribute to heating the primary (proton,
electron, helium) plasma in the extended corona and accelerating
the solar wind?}
CPI will determine the degree of bulk-plasma heating and acceleration
from the most likely physical processes.
The properties of the resonant wave spectrum (from the previous question)
are key inputs to this question as well, but additional
CPI observations are needed to constrain the
full range of processes.
The primary UV measurements for this question are the proton, electron,
and helium plasma properties; e.g., H~I Ly$\alpha$
resonance scattering line profiles (H$^{0}$ is
closely coupled to protons by charge transfer;
Allen et al.\  2000), H~I Ly$\alpha$ Thomson scattering
line profiles (a probe of the electron velocity
distribution), and He~II 30.4 nm line profiles.
The methodology for this question involves comparing heating and
acceleration rates (derived from measured temperatures, densities,
and outflow speeds) to rates predicted by theoretical models of
the following:

\vspace*{0.03in}
\noindent
\newcounter{bean}
\begin{list}{\arabic{bean}.}{\usecounter{bean}%
\setlength{\leftmargin}{0.17in}%
\setlength{\rightmargin}{0.0in}%
\setlength{\labelwidth}{0.17in}%
\setlength{\labelsep}{0.05in}%
\setlength{\listparindent}{0.0in}%
\setlength{\itemsep}{0.03in}%
\setlength{\parsep}{0.0in}%
\setlength{\topsep}{0.0in}}

\item[1.]
{\bf High-frequency (or other ion-resonant kinetic) waves:}
The information about the resonant
wave spectrum obtained from minor ions
can be extrapolated to predict proton and
helium energization rates that are associated
with the specific ``flavor'' of wave-particle interaction
that is evaluated to be consistent with
the ion measurements discussed above.

\item[2.]
{\bf Low-frequency waves:}
CPI can constrain the amplitudes of low-frequency waves that
may not be efficient at heating minor ions. These
waves can accelerate the plasma by wave pressure
and possibly heat electrons via Landau damping.
Preliminary observations of long-period (10--20 minute) compressive
fluctuations in coronal holes have been made (e.g.,
DeForest \& Gurman 1998; Krishna Prasad et al.\  2011),
but CPI can confirm and better characterize the full plasma response
to these fluctuations.
Also, models that predict significant ponderomotive acceleration
from low-frequency waves (Ofman \& Davila 2001) can be tested by
constraining the available power.

\item[3.]
{\bf Suprathermal particles:}
Ubiquitous in space plasmas, particles in the tails of proton
and electron velocity distributions have been
suggested as key contributors to coronal heating
(Scudder 1992; Vi\~{n}as et al.\  2000) and in
determining {\em in~situ} charge states (Esser \& Edgar 2000).
The presence of strong tails in the very low corona is doubtful
because of the high collision rate (Feldman et al.\  2007),
but CPI can detect these tails in proton 
and electron distributions (via departures from Gaussian
shapes in the H~I Ly$\alpha$ profiles) well into the collisionless
domain at large heights.

\end{list}

\vspace*{0.05in}
\noindent
CPI will also answer detailed questions about mapping the origins
of {\em in~situ} solar wind streams down to the coronal magnetic field,
and about the role of episodic jets and reconnection in the
solar wind's mass and energy budget.
These questions are discussed further in Section 2.3 below, along
with the contributions of other NASA heliophysics missions.
In addition, the improved understanding of coronal physics that CPI will
facilitate is important not only for explaining the origins of
space weather, but also for establishing a baseline of knowledge
directly relevant to other stars and astrophysical systems
ranging from the interstellar medium to black hole accretion disks.

\vspace*{0.05in}
\begin{center}
{\em 2.2. Coronal Mass Ejections (CMEs)}
\end{center}

\vspace*{-0.01in}
CMEs are a secondary objective for CPI because solar minimum conditions
will prevail during the mission's primary phase.
Nevertheless, there will be one CME
every 2--3 days based on LASCO observations
(Yashiro et al.\  2004), and we expect to observe
30 to 50 CMEs during the primary mission and
considerably more during the extended mission
thanks to increasing solar activity.
UVCS demonstrated the power of spectroscopy
to unravel the physics of CMEs (see
Kohl et al.\  2006), but new observational capabilities
are needed to answer the most fundamental questions.
CPI will provide a wide range of unambiguous diagnostics for the
physical conditions of CMEs.
This is enabled by a multi-slit capability and the sensitivity
and spectral range that permit observation of
He~II 30.4 nm and several density-diagnostic line ratios
(e.g., Ne VII 89.5/88.7 nm and O VI 55.4/55.5 nm) to
constrain time-dependent ionization models of the expanding plasma.
CPI will address the following questions.

{\em What physical processes drive the evolution and heating
of CMEs in the corona?}
UVCS data have been used to reconstruct the 3D
structures and thermal histories of a small
number of CMEs, and there is evidence that the plasma in CMEs
is strongly heated as it accelerates
(e.g., Lee et al.\  2009; Landi et al.\  2010).
The nature of the heating is not known, but a number of
energization mechanisms have been proposed
(Kumar \& Rust 1996; Lin et al.\  2004; Reeves et al.\  2010;
Murphy et al.\  2011).
CPI will help distinguish between the predictions of these
different models by measuring plasma parameters not only in the
brightest parts of CMEs (e.g., the
bright front and the twisted remnants of the
prominence or flux rope) but also in the dimmer ``void''
region behind the front and the shock ahead of the front.

{\em What is the role of magnetic reconnection 
in the eruption and relaxation of the 
coronal magnetic field in CMEs?}
In the classic CME picture (Sturrock \& Coppi 1966),
an erupting flux rope leaves behind
oppositely directed open fields that are
separated by a current sheet.
Reconnection in this sheet produces an arcade of post-CME
loops seen as large X-ray flares,
at the same time allowing the twisted flux rope to escape
as the CME plasmoid (Lin \& Forbes 2000).
In a few events,
UVCS detected spatially narrow emission features in [Fe XVIII] 97.4 nm
(formed at $T_{e} \approx 6$ MK) that match the expected
properties of thin current sheets (Raymond 2008).
CPI will observe many more lines from the
extremely hot plasma at these reconnection
sites, up to [Fe XXII] 82.2 nm, with improved sensitivity.
Line ratios will be used to infer the electron temperature,
density, and flow speeds (Ko et al.\  2010).
Line widths measured by CPI can be used to put limits on the
velocity amplitude of turbulent motions in current sheets
(see Kowal et al.\  2009).

\vspace*{0.02in}
\begin{center}
{\em 2.3. Relationships to Other NASA Missions}
\end{center}

\vspace*{0.01in}
A unique strength of CPI is how it both strongly complements and
is complemented by other NASA solar missions planned
for the same time period.
The following questions represent examples of the wider science
objectives achievable by integrating CPI measurements
with those from other missions.

{\em How do nanoflares and magnetic reconnection affect
the solar wind?}
UV and X-ray imaging instruments on {\em SDO, Hinode, STEREO, IRIS,}
and {\em Solar Orbiter} will put tight new constraints
on the mechanisms responsible for the intermittent energy
release from reconnection events below $\sim$1.2 $R_{\odot}$.
CPI measurements above 1.4 $R_{\odot}$ provide the
radial dependence of the heating rates and constrain
the parameters of reconnection-based solar wind models
(Fisk 2003; Antiochos et al.\  2011).
CPI will also have the sensitivity to identify rapid thermal changes
associated with the postulated field-line motions in jet-like
reconnection events that are detected closer to the disk with
instruments like AIA/{\em{SDO}} and {\em IRIS} (e.g.,
Yang et al.\  2011).

{\em How do plasma properties and magnetic field lines connect
from the photosphere to the heliosphere?}
A key goal of space weather prediction is the ability to map
magnetic field from the Sun to 1 AU.
Currently this is done by using disk images and magnetograms
(e.g., MDI/{\em{SOHO}} and HMI/{\em{SDO}}) as lower boundary
conditions and {\em in~situ} particle and field data to test the
results of MHD models.
CPI makes possible a more complete picture of this 3D system by providing
the ``missing link'' between data at 1 $R_{\odot}$ and 1 AU.
It also aids in the interpretation of inner heliospheric measurements by
{\em Solar Probe Plus} and {\em Solar Orbiter} by probing the
plasma properties in the wind's acceleration region.
The stable coronal structure at solar minimum will permit tomographic
reconstruction of these plasma properties (Panasyuk et al.\  1999;
Frazin et al.\  2009; van der Holst et al.\  2010), so that global
MHD models can be better tested and refined.
CPI also provides elemental abundance and charge state data
for minor ions, which is another way to trace the origin of
{\em in~situ} solar wind streams (e.g., Ko et al.\  2006; Zurbuchen 2007).
CPI will build up an unprecedented
database of abundance maps that will test models of
fractionation and the First Ionization Potential (FIP) effect
from gravitational settling (Lenz 2004) and MHD waves (Laming 2009).

\vspace*{0.05in}
\begin{center}
{\bf 3. The CPI Science Payload}
\end{center}

\vspace*{-0.15in}
\begin{center}
{\em 3.1. Instrumentation Overview}
\end{center}

\begin{figure}[!t]
\vspace*{-0.02in}
\hspace*{0.02in}
\epsfig{figure=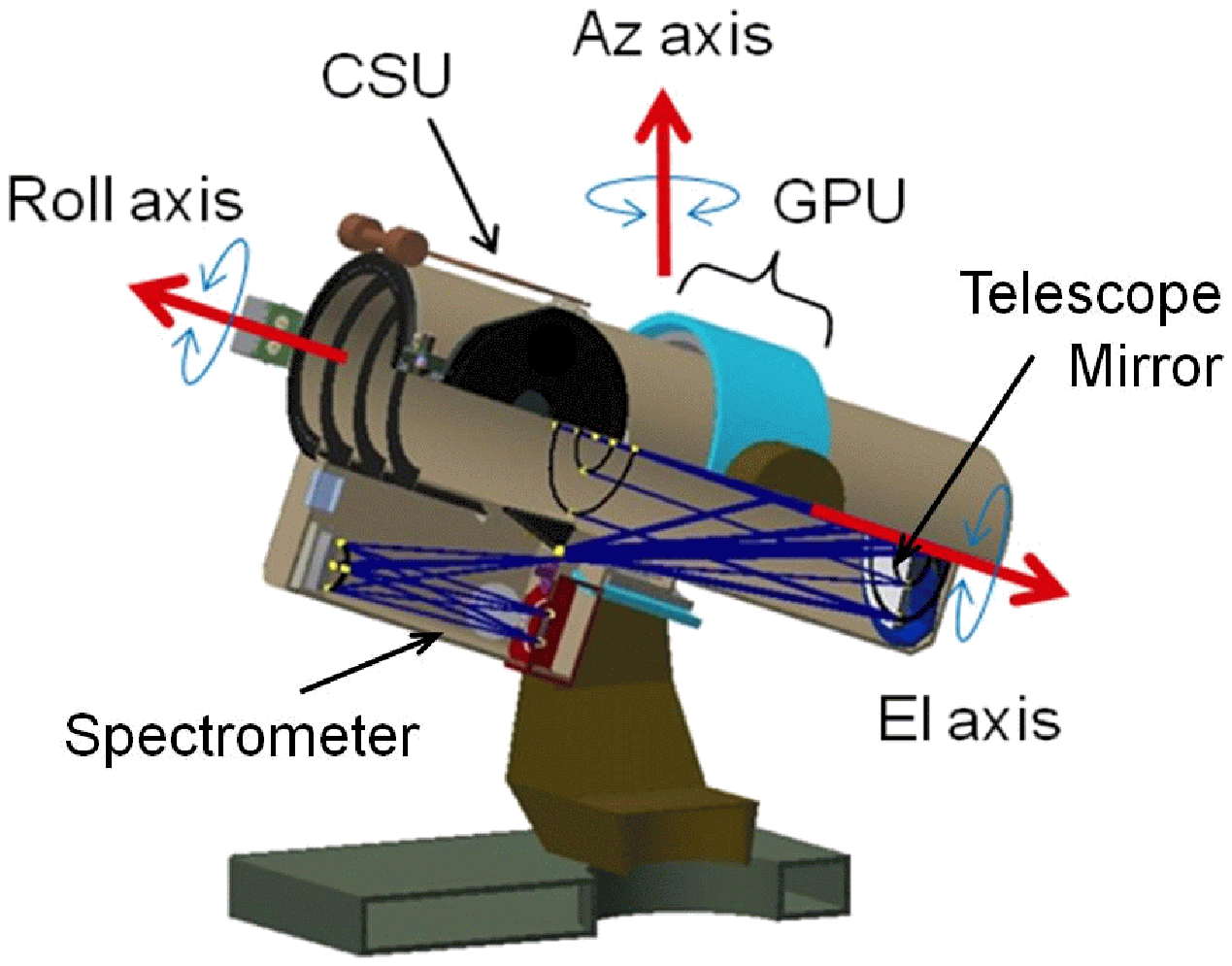,height=2.60in}

\vspace*{-2.61in}
\hspace*{3.43in}
\epsfig{figure=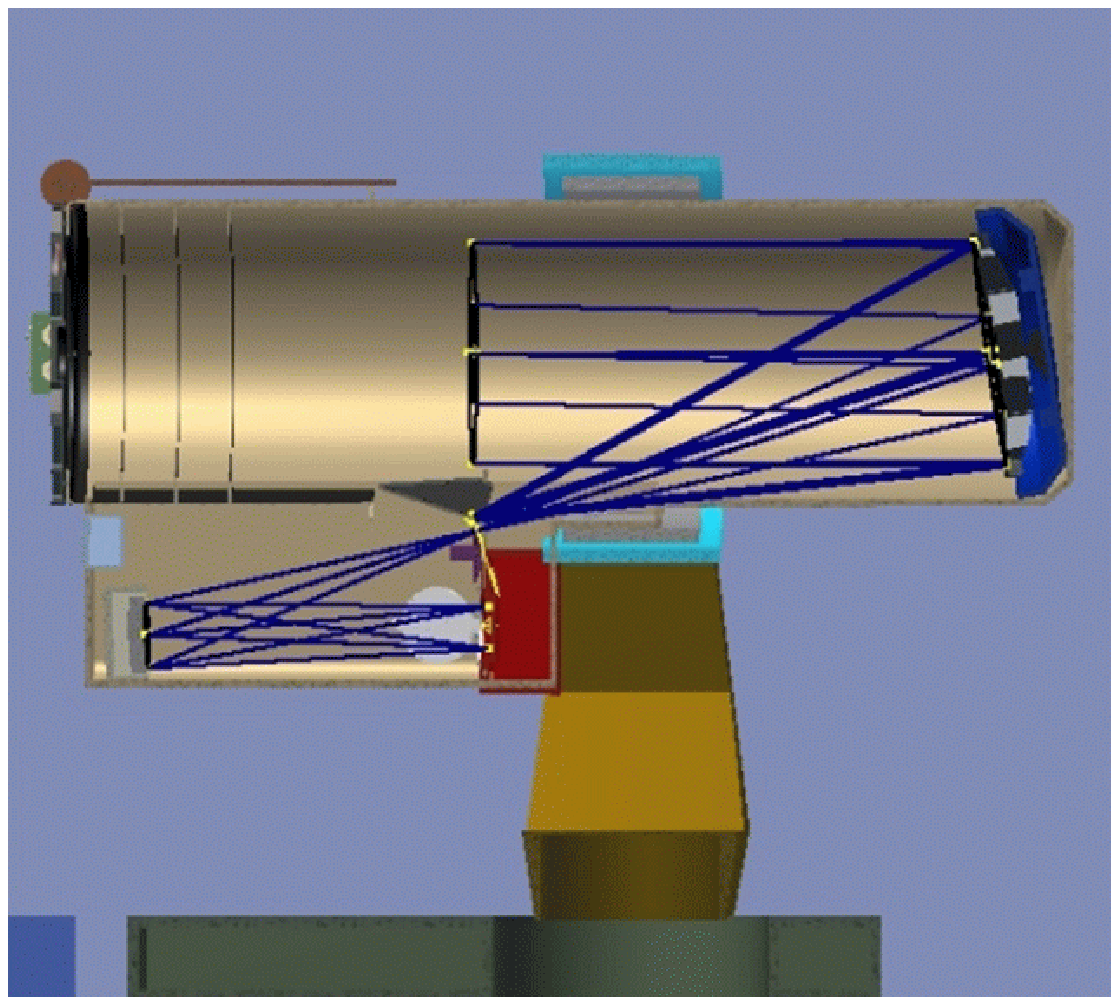,height=2.60in}

\vspace*{0.07in}
\hspace*{1.22in}
\small
{\bf Figure 3:}
CPI in its on-orbit (left) and transport (right) configurations.
\end{figure}
The CPI Science Payload (SP) is comprised of the 
Coronagraph-Spectrometer Unit (CSU), the Gimbals Pointing Unit (GPU), 
and the Remote Electronics Unit (REU); see Figure 3.
The CSU is further comprised of the Telescope Assembly (TA) and
the Spectrometer Assembly (SPA).
The TA feeds light to the SPA, which consists of a toric dual-grating
spectrometer with an intensified CCD (ICCD) detection system.
The SP has two optical paths: the EUV path (EUVP) for spectroscopy
at 68.0--126 nm,
and the He~II path (HeP) for spectroscopy at 49.5--84 nm 
(1st order) and 25.7--42.0 nm (2nd order).
It obtains high resolution spectra of spatial regions as small as 4.5$''$.
The field of view is shown in Figure 4.
\begin{figure}[!t]
\vspace*{-0.01in}
\hspace*{1.40in}
\epsfig{figure=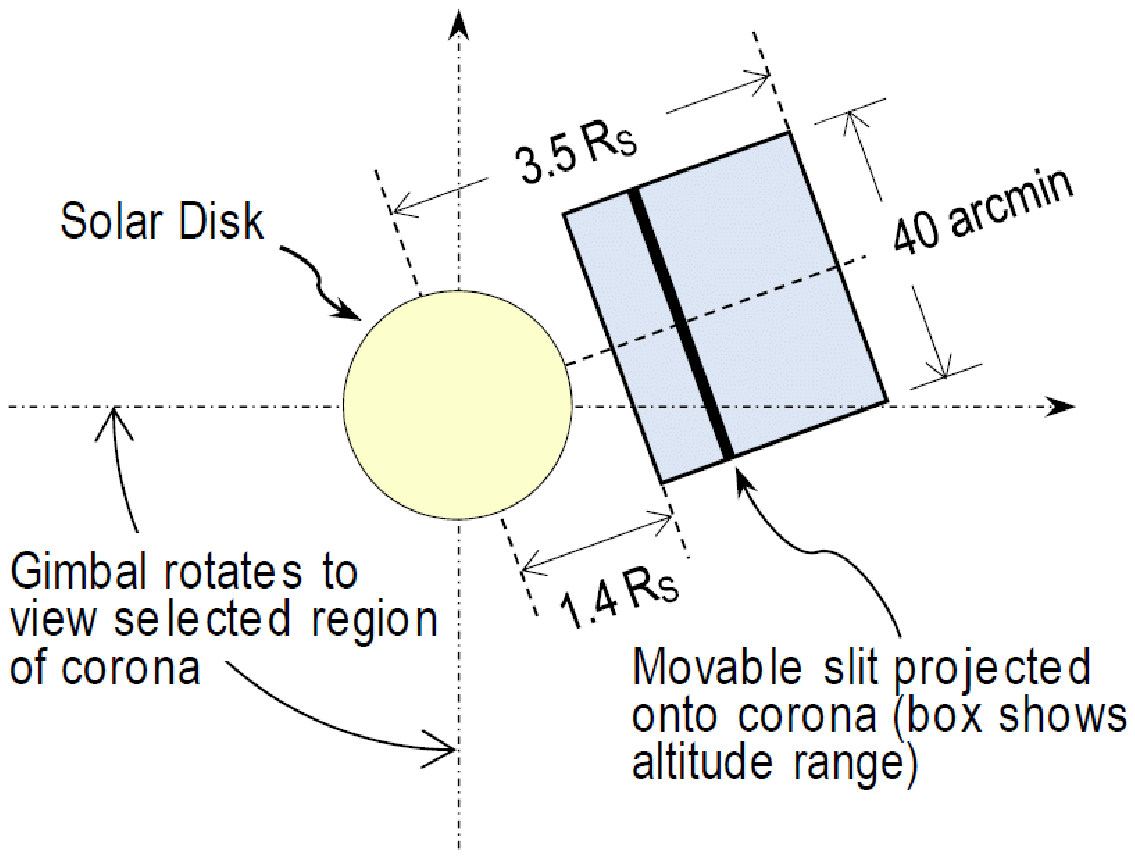,width=3.70in}

\hspace*{2.10in}
\small
{\bf Figure 4:}
The CPI field of view (FOV).

\vspace*{0.15in}
\hspace*{1.25in}
\epsfig{figure=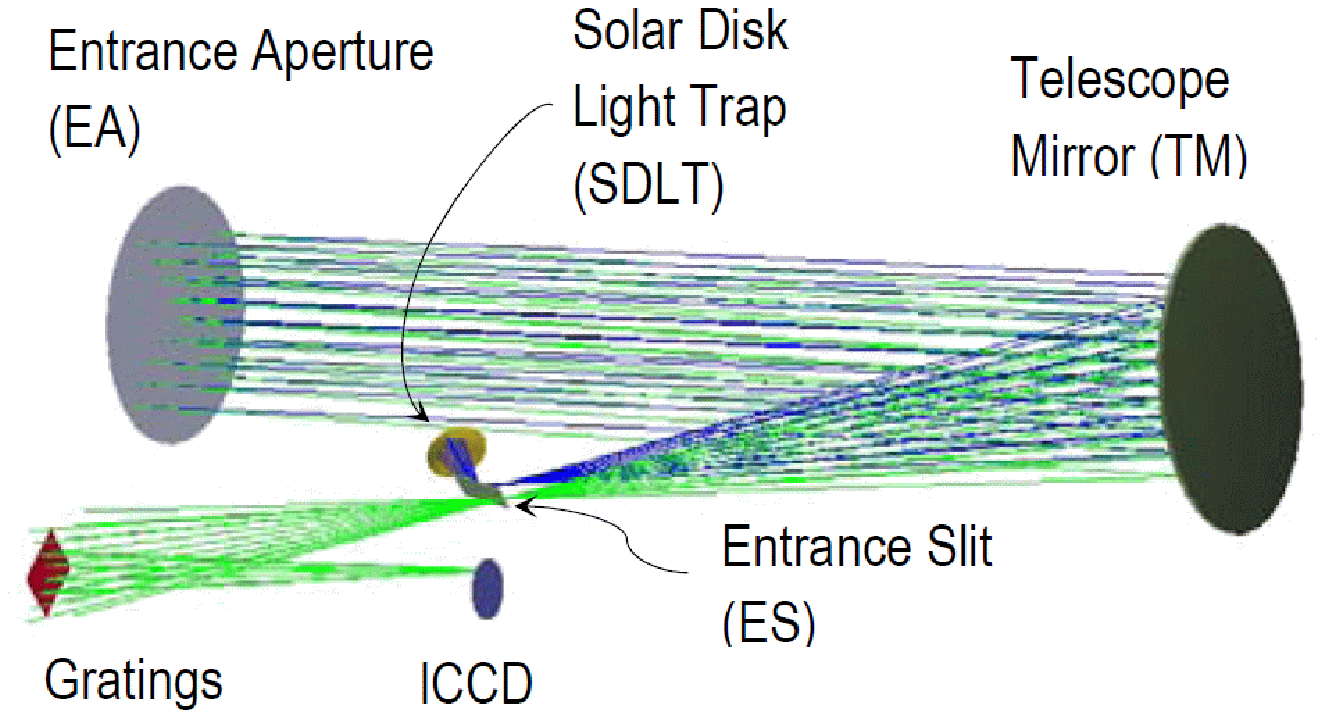,width=4.00in}

\vspace*{0.04in}
\small
{\bf Figure 5:}
ZEMAX rendering of CPI optical layout.
Shown are coronal rays (green) from an altitude of
3.5 $R_{\odot}$ that pass into the spectrometer, while disk
light (blue) is redirected into the Solar Disk Light Trap.
Rays that miss the grating include all rays from the EA.
\end{figure}

A key capability of CPI is that it facilitates determination and removal
of the stray light spectrum for each observed height with solar disk 
observations scaled with the intensities of isolated stray light 
spectral features observed during coronal measurements.

The CSU optical layout is shown in Figure 5.
Light enters through the 16.8 cm diameter Entrance Aperture 
(EA) and falls on the Telescope Mirror (TM), which is under-filled for
both the solar disk and coronal radiation.
One half of the mirror surface is SiC for illumination
of the EUV path and the other half is 
coated with Si/Mo-Ir for the He~II path.
For coronal observations, the GPU and TM place an image of
the disk on the Solar Disk Light Trap, which is
elongated in the radial direction to allow coronal images at heights
of 1.4 to 3.5 $R_{\odot}$ to be centered on the SPA Entrance Slit.
This light trap is similar to that of UVCS/{\em{Spartan}}
(Kohl et al.\  1994, 2006) and consists of a polished
high-reflectance mirror at an oblique angle that reflects
the direct solar disk radiation into a blackened cavity with
close-packed 5 mm diameter tubes.
The mirror forms a 1.7 times magnified image of the EA at 136 cm,
which is beyond the SPA Entrance Slit and Gratings.
Most of this light is intercepted by the slit jaws, and the portion that
passes through the slit is intercepted by a light trap that is located 
just in front of the gratings and surrounds them.
There are baffles between the entrance aperture and telescope mirror,
and all non-optical surfaces in the telescope are designed to
attenuate stray light.
There is no CPI hardware Sunward of the EA.

The Toric Dual-Grating Spectrometer has an Entrance Slit mechanism
with 5 selectable slits of varying widths and a set of three slits
for simultaneous observations at multiple heights.
Coronal light passes through the slit and onto two toric gratings,
one for each light path. 
The gratings act nearly as one optic.
The grating size is chosen to not accept any light that
contributes to the image of the Entrance Aperture including
diffraction of that light by the slit,
as well as to not accept light from the outer
portion of the mirror that would have unacceptable imaging properties.
The gratings are slightly tilted with respect to 
one another in the perpendicular to dispersion direction, so that their 
focal planes are offset by 2.0 mm at the detector.
They are oriented in the dispersion direction to minimize overlap
between the coronal lines of the two paths.
There is a grating drive that is used to optimize 
image/spectral quality for spectral lines near the outer limits of the 
EUV wavelength range, and also to facilitate avoidance of any detector 
imperfections.
The pivot axis is 4.09 cm from the center of the grating surface.
The SPA has light traps for zero order and for intercepting 
light from the Entrance Aperture edges as well as a baffle located 
between the gratings and the detector to ensure that the detector only 
views rays from the gratings.

\begin{figure}[!t]
\vspace*{-0.01in}
\hspace*{1.00in}
\epsfig{figure=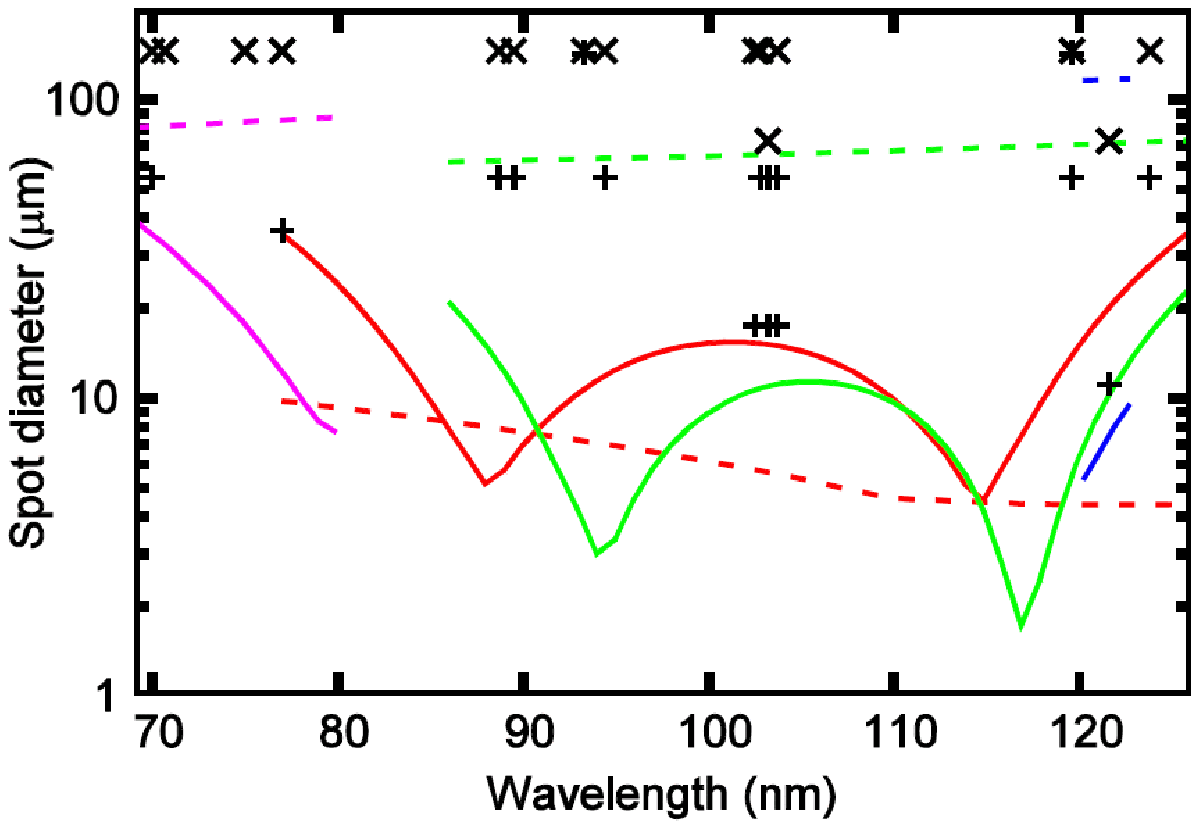,width=4.40in}

\vspace*{0.04in}
\small
{\bf Figure 6:}
Predicted spectral (solid lines) and spatial
(dashed lines) FWHM spot diameters of the EUV channel.
Grating positions for four observation modes are shown:
standard (red), jet observations (green),
H~I Ly$\alpha$ (blue), and short wavelength (pink).
Points represent spectral ($+$) and spatial ($\times$) requirements.
\end{figure}
Optical performance of CPI was modeled with ZEMAX and is illustrated 
for the EUV path in Figure~6.
The HeP performance also meets all of the requirements.
CPI provides a tremendous improvement in count rates over 
UVCS/{\em{SOHO}} for the portion of the CPI wavelength range covered
by UVCS (e.g., factors of 200, 40, and 20 at 1.4, 1.8, and 2.5
$R_{\odot}$, respectively for observations of O~VI 103.2 nm).
This is the result of a greatly increased unvignetted telescope
area and improved reflective coatings and detector counting efficiency.

\newpage
\begin{center}
{\em 3.2. Stray Light Suppression}
\end{center}

Stray light suppression is sufficient for all required observations 
(see Figure 7).
The fundamental and dominant source of stray light is due to
diffraction by the finite size of
the entrance aperture and telescope mirror.
To minimize this diffraction, the entrance aperture and mirror
are oversized so that the focal ratio is much smaller ($f$/2.5)
than the focal ratio of the gratings (effective focal ratio of $f$/7).
In this way, the diffracted light from the mirror that passes
through the SPA slit and
onto the gratings can be reduced approximately in proportion to
the diameter of the entrance aperture. 
Other sources of stray light are made negligible with careful design.
Direct solar-disk light is specularly reflected off the telescope 
mirror and imaged onto the Solar Disk Light Trap, which reduces it to 
negligible levels.
Solar disk radiation that encounters particulate contamination on
the reflecting surface of the telescope primary has been shown to
be controllable to negligible levels.
Scattering due to mirror figure errors and microroughness are
reduced to near negligible levels.
Observational lines of sight are restricted such that glints off
surfaces of the ISS illuminate the EA at large ($> 10^{\circ}$) angles.
This prevents any glint light from falling on the portion of the telescope 
mirror that has straight-line or slit-diffracted paths to the gratings. 
Baffles on the TA structure effectively eliminate glints that enter the 
EA at angles that miss the mirror.
It is also necessary to reduce the solar disk light at wavelengths
outside the band pass of the spectrometer.
This off-band radiation is eliminated by the diffraction 
gratings for wavelengths shortward of 180 nm.
Suppression at longer wavelengths, where the solar disk intensity
is extremely high, is provided by a combination of the occulting
system, the diffraction grating, and the low sensitivity of
the ICCD detector.  
\begin{figure}[!t]
\vspace*{-0.01in}
\hspace*{1.08in}
\epsfig{figure=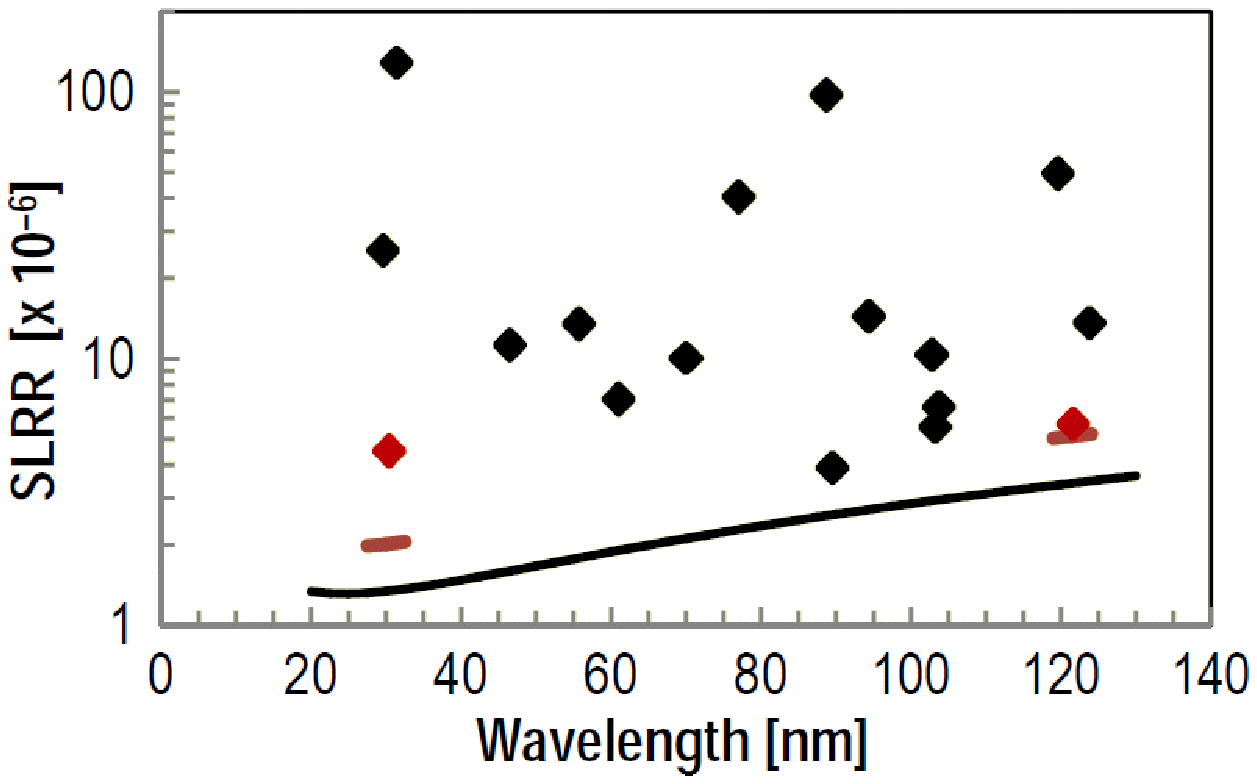,width=4.20in}

\vspace*{0.04in}
\small
{\bf Figure 7:}
Stray light reduction ratios. Lines
indicate modeled CPI stray light levels at 1.8 $R_{\odot}$
(red: He~II 30.4 nm, H~I Ly$\alpha$) and 2.0 $R_{\odot}$
(black: all other lines).
Corresponding colored points show stray light requirement
for coronal hole spectral lines.
\end{figure}

Stray light characterization and removal will be accomplished with a
combination of a stray light model together with laboratory stray light 
characterization, and during the mission, observations of the solar 
disk spectrum and direct detection of spectral features known to be due 
solely to stray light.
The model will be refined during the instrument development.
At this time, the stray light levels have been computed 
with a series of empirically validated models that verified that the 
dominate contributor is finite-aperture diffraction from the entrance 
aperture.
The other possible sources of stray light were found to be
controllable and can be made negligible.
A simple parameterized model of non-specular stray light was
used to predict the expected stray light levels.
The results of this analysis are provided in Figure 7. 
This model, which was validated by adapting it to the geometry and 
optical properties of UVCS/{\em{SOHO}}, predicts a level of stray light
that agrees with in-flight measurements (see Cranmer et al.\  2010).
The model was used to determine preliminary microroughness and
figure error specifications for the TM.

Modeling indicates that for most spectral lines of interest, the ratio 
of the coronal signal to the stray light will be sufficient to achieve 
the science goals with a minimum background analysis.
However, for some important lines, CPI will be required to accept
and remove stray light levels that are comparatively high;
for example, reaching an estimated level of 1:1.5 corona-to-stray
light ratio at 1.8 $R_{\odot}$ for H~I Ly$\alpha$ and He~II 30.4 nm
in coronal holes.

The characterization and removal of the stray light background 
will use a combination of the synchronous solar disk observations (with 
a reduced aperture to bring the count rate within the dynamic range of 
the ICCD detector), the measurements of spectral features produced 
solely by stray light, and coronal hole observations above 2.5 $R_{\odot}$, 
which will be almost entirely stray light in the case of
H~I Ly$\alpha$ and He~II 30.4 nm.
Laboratory stray light calibration data will be used to 
extrapolate stray light contributions to lower altitudes, and to 
determine the stray light level as a function of the origin of the 
light on the disk and the observed height above the solar limb in order 
to verify parameters in the stray light model, which in turn will be 
used with in-flight solar disk observations to weight the stray light 
contributions from various parts of the disk.
All observations will be modeled as a combination of coronal emission,
stray light, and detector background.
In addition, H~I Ly$\alpha$ observations will include a model 
of geocoronal emission and absorption.
All the components except coronal emission have well known
or separately measurable spectral shapes.
The relative contributions of the components will be varied to 
achieve the best fit to the observed data, but the range of their 
variation will be highly constrained by the laboratory stray light 
model, the measured stray light spectral features, and the solar disk 
spectrum (as well as, in the case of H~I Ly$\alpha$, a geocoronal model 
similar to the one used for the UVCS/{\em{Spartan}} (Kohl et al.\  1996). 

Coronal emission will be modeled as a combination of a Gaussian core 
and high-velocity wings, corresponding to electron scattering and 
possibly suprathermal particle scattering.
The characteristic spectral widths of the coronal emission
components are significantly higher than the rest of the
contributing features, which facilitates high-precision 
component separation.
In summary, because (1) a high quality advance knowledge
of the stray light characterization will exist, (2) measurements
of the stray light will exist at nearby wavelengths, and (3)
very recent measurements of the solar disk spectrum with the same 
instruments will exist, then model fitting with a very limited number of
highly constrained ``free parameters'' will result in the stray light 
removal procedure being extremely robust.
The precision of the coronal profile recovery is expected to be
at the level of 5\% or better, even for signal to stray light
ratios of greater than 1:1.5.

\vspace*{0.02in}
\begin{center}
{\em 3.3. ICCD Detector}
\end{center}

\vspace*{0.01in}
The detector consists of a 40~mm diameter microchannel plate (MCP)
intensifier with a fiber optic output window optically coupled to a CCD.
The MCP is coated with KBr.
The intensifier includes a vacuum door and non-flight 2 liter/s ion pump.
This detector has a high technical reliability since
it is essentially the same technology that the 
Mullard Space Science Laboratory (MSSL) used for the ICCD detectors on 
{\em Swift} (Roming et al.\  2000) and {\em XMM/Newton}
(Mason et al.\  2001). 
Following assembly, the detector is maintained under vacuum to prevent 
deterioration of the photocathode.
A grid/aperture in front of the MCP repels low energy ions.
The grid is set to $\sim$100 V, and is adjustable in flight. 

Individual photoelectrons are produced at the MCP surface and 
multiplied through its pores. Exiting electrons are accelerated onto a
phosphor screen coupled to a CCD chip using a 3.5:1 fiber optic taper. 
Each photoelectron results in a ``splash'' of light on a
$3 \time 3$ pixel area of the CCD.
The light splash is centroided to 0.25 pixels (spectral only),
improving the resolution. A $1024 \times 256$ window is read out, 
allowing an increased frame rate. 

If two events land in the same $3 \times 3$ pixel area during an
exposure, the two events can be identified using a combination
of pulse height and position information, and are both
rejected from the image (although retained in a count register).
To minimize this effect, the CCD is 
clocked at the maximum frame rate.
The maximum acceptable count rate for CPI of 5.6 counts/s at
a 50\% loss rate is somewhat increased by two factors:

\vspace*{0.03in}
\noindent
\begin{list}{\arabic{bean}.}{\usecounter{bean}%
\setlength{\leftmargin}{0.17in}%
\setlength{\rightmargin}{0.0in}%
\setlength{\labelwidth}{0.17in}%
\setlength{\labelsep}{0.05in}%
\setlength{\listparindent}{0.0in}%
\setlength{\itemsep}{0.03in}%
\setlength{\parsep}{0.0in}%
\setlength{\topsep}{0.0in}}

\item[a.]
The position resolution required in the spatial dimension is less 
critical than in the spectral dimension. This gives some relaxation in 
the proximity of events in that axis, where centroiding is not required.

\item[b.]
The maximum count rates are at the peaks of spectral lines. The 
shoulders and base of the lines are within 3 CCD pixels, so that the 
average count rate over 3 pixels is lower than at the peak. This effect 
allows a higher maximum count rate for the disk observations compared 
to the coronal observations: lines on the disk are narrower, and so a 
higher peak count rate can be tolerated because the adjacent areas 
within the 3-CCD-pixel width have a lower rate. 

\end{list}

\vspace*{0.02in}
\begin{center}
{\em 3.4. Line of Sight Pointing System (LOSPS)}
\end{center}

\vspace*{0.01in}
The tracking and pointing of the 
CPI Coronagraph Spectrometer Unit is provided by the Gimbals Pointing 
Unit (GPU) and a Telescope Mirror Pointing Subassembly (MPS). The GPU 
provides the coarse tracking of the Sun across the sky during orbital 
day, and the fine pointing of the instrument so that the solar disk image 
falls on the Solar Disk Light Trap and the coronal region of interest 
falls on the spectrometer entrance slit.
The CSU is mounted in a Roll Axis Assembly that provides
rotation of the CPI field of view about the 
Sun-center axis to place the image of the selected position angle axis 
on the center of the spectrometer slit.
The Roll Axis Assembly is mounted in a yoke that provides rotation in
the elevation axis, and the yoke is mounted on an Azimuth Assembly
that provides rotation about the axis that is perpendicular to the
elevation axis.
Low frequency disturbances are 
removed by the GPU. Higher frequency vibrations over the small 
amplitudes expected on ISS are compensated with motions of the 
telescope mirror provided by the MPS. L-3 Com Integrated Optical 
Systems is highly experienced in such systems, having developed similar 
ones for the {\em WISE, HIRDLS, GIFTS,} and {\em GOES-R ABI}
missions.

\vspace*{0.05in}
\begin{center}
{\bf 4. Science Payload Accommodation on ISS}
\end{center}

ISS engineers at JSC have provided a preliminary confirmation that the
CPI accommodation requirements are compatible with the ISS. 

\vspace*{0.02in}
\begin{center}
{\em 4.1. ISS External Research Accommodation Sites Characteristics}
\end{center}

\vspace*{0.01in}
The primary ISS site for CPI is the Columbus Starboard Overhead
Zenith (SOZ) site. This is the 
Columbus External Payload Facility zenith position which has been used 
for the Columbus SOLAR payload. The CPI accommodation requirements are 
well within the allowable SP mass, power, SP volume, and telemetry rates 
specifications for this site.
Mounting, power, and data interfaces are standardized on
pallet plates that ease mounting, robotic handling, and
installation (e.g., the Columbus standard cargo interface called the
Flight Releasable Attachment Mechanism; FRAM). 
After installation, the ISS provides data up/downlink and command and 
control via a standard interface at MSFC.
At the end of the instrument's lifetime, an established
ISS procedure removes the payload and transports 
it for safe discard.

\vspace*{0.02in}
\begin{center}
{\em 4.2. CPI ISS Resource Requirements}
\end{center}

\vspace*{0.01in}
ISS capability provides a 54\% reserve in addition to the CPI
requirement for mass and 376\% for power.
CPI fits in the Columbus SOZ site and transport volume with
ample reserves.
The CPI data rate and data capacity requirements are less than 2\%
of the ISS capable data rate for both S and K band transmitter systems.
Thermal ISS interfaces are capable of supporting CPI requirements.
ISS contamination experts have informed us that the ISS is
compatible with the CPI particulate and molecular 
contamination requirements as long as CPI follows its plan to never 
point at an ISS structure and to always close its Entrance Aperture 
Door during high contamination periods, 
which occur less than a few percent of the time.

\vspace*{0.02in}
\begin{center}
{\em 4.3.  ISS Orbit and Trajectories}
\end{center}

\vspace*{0.01in}
The ISS ram direction is the $x$ axis, the 
zenith is the $z$ axis, and the orthogonal axis is $y$.
The current ISS altitude varies from 350 to 460 km and is
expected to increase to an average of 450 km prior to CPI operations.
Atmospheric absorption will allow scientific observations to
commence before the Sun vector reaches the $xy$ plane.
After consultation with the ISS MAGIK team, we calculate 
a typical minimum solar viewing time at SOZ of 35 minutes per orbit
for at least 85\% of the orbits plus 5 minutes for acquisition
during the 19$^{\circ}$ below the $xy$ plane.
These calculations account for a 10$^{\circ}$ unobstructed 
exclusion window about our line of sight (LOS).
Observations begin when the Sun angle reaches the ISS $xy$ plane
and continue past the ISS $yz$ plane for a total angle of
$\sim$137$^{\circ}$.
The $\beta$ angle is defined as the 
angle between the orbit plane (ISS $xz$ plane) and the vector from the 
Sun (e.g., $\beta = 0$ when the Sun is in the ISS $xz$ plane and
$+$90$^{\circ}$ when it is in the ISS $yz$ plane on the starboard side).
For $\beta$ from 0 to $+$40$^{\circ}$ 
our viewing time is $\sim$40 minutes and for 0 to --40$^{\circ}$ 
it is $\sim$35 minutes due to a truss obstruction on the
port side and an ISS module structure to the rear.  
At $\beta$ angles of $\pm 55^{\circ}$, we are limited by solar
array obstructions and so there will be no CPI observations
at $\beta$ exceeding these angles.
Preliminary review of the JEM site \#3 indicates that it is nearly 
equivalent to SOZ.

\vspace*{0.05in}
\begin{center}
{\bf 5. Mission Overview}
\end{center}

\vspace*{-0.15in}
\begin{center}
{\em 5.1.  Mission Design}
\end{center}

\vspace*{-0.01in}
CPI is an Explorer Small Complete Mission intended for 
the International Space Station (ISS).
The baseline ISS site is the Columbus EPF
Starboard Overhead Zenith (SOZ) site, and the alternate 
site is JEM-EF EFU \#3.
CPI utilizes the ISS resources for transport to 
the site, attachment via the robotic arm, and for ISS provided 
telemetry links. The nominal launch date is 1 September 2017.
On-orbit checkout continues for two months, followed by
24 months of coronal and solar disk observations.
CPI will acquire the Sun at the start of 
orbital day, starting observations after 5 minutes, and continuing them 
for about 35 minutes or more.
The current average orbital altitude is 400 km and is
expected to have increased at the time of CPI operations after 
the NASA Space Transportation System is retired.
The Sun trajectory rises rapidly in the sky during orbital day,
providing minimal atmospheric absorption during the observation period.
It is estimated that ISS will provide 310 days per year with
16 orbits per day and at least 35 minutes of solar observations
(not including 5 minutes for Sun acquisition) per orbit.
ISS is well suited to solar experiments because the
solar panels are normally aimed at the Sun and out of the line of 
sight of solar-viewing instruments. ISS engineers assure us that the 
contamination environment meets the stringent requirements of CPI, but 
precautions must be taken to not point CPI at the ISS surfaces, and to 
close its aperture door at times of 
contamination release and the presence of transport vehicles.
A three-year extended mission is proposed. The entire SP
will be designed to meet ISS hardware disposal requirements.

\newpage
\begin{center}
{\em 5.2. Operations, Data Policy, and Data Archiving}
\end{center}

\vspace*{-0.02in}
The Science Operations Center will be located at the Smithsonian
Astrophysical Observatory (SAO).
Ground software will be provided by SAO and based on the command
software it developed and used for UVCS/{\em{SOHO.}}
The command and telemetry link to the ISS will be provided
by the ISS Payload Operations Center (POC) at MSFC.
The SAO command software will be upgraded to support 
communication protocols of the MSFC command and telemetry
software that is to be provided to SAO. 

CPI will follow NASA's Heliophysics Science Data 
Management Policy guidelines to make the science data publicly 
available as quickly as possible.
Validated science data will be distributed via the
Internet directly from SAO and through the Virtual Solar Observatory
with assistance from GSFC's SDA.
Level~1 data will be distributed within one month of the time
when the data are available to SAO.
This will ensure wide dissemination of all
CPI data products to the science community. 

The validated CPI data products---including associated
documentation on calibration methods---will be submitted to 
the appropriate NSSDC archive as they become available.
Level 2 data, calibrated with the best and final calibration
parameters, will be provided during the last month of the
primary mission.

\vspace*{0.02in}
\begin{center}
{\em 5.3. Guest Investigator Program}
\end{center}

\vspace*{-0.02in}
The CPI science team will support an active Guest Investigator
(GI) Program during both the primary baseline mission and its
continuation to the extent desired by NASA/SMD.
It is envisioned that an announcement of opportunity will be
released shortly after launch that would consider a broad range
of guest participation in the CPI Investigation, supported by
about 15 grants per year.
These programs will be selected via a rigorous NASA peer
review process with the goal of broadly complementing the
core team science.
GIs will be treated exactly as CPI Co-Investigators, including
any desired participation in observation planning.

\vspace*{0.05in}
\begin{center}
{\bf 6. The CPI Team, Management, and Costs}
\end{center}

\vspace*{-0.13in}
\begin{center}
{\em 6.1. The CPI Team}
\end{center}

\vspace*{-0.02in}
The CPI Project is a collaborative effort under the direction of John 
Kohl, the Principal Investigator, involving a core scientific team from 
SAO, the University of Montana, and the University of New Hampshire.
The PI has delegated the authority to manage the CPI project to the
Project Manager, Timothy Norton.
The Mission Systems Engineer is Peter Daigneau.
George Nystrom will lead the project-level reviews team.
The industrial partner is L-3 Communications Integrated Optical Systems 
(L-3 LOS).
Paul Cucchiaro will be the Payload Project Manager for L-3 Com,
and Mark Schwalm will be the Payload Systems Engineer. L-3 LOS and 
its collaborating L-3 Com organizations are highly experienced in the 
development of space instrumentation and ISS systems and integration,
with over 30 years of experience and a 100\% success rate on 50$+$ space 
flight payloads.
It has developed and fielded space-based gimbals, 
pointing mirrors, SiC UV optics, and grating spectrometers, providing 
high technical readiness levels against the CPI payload needs.
L-3 LOS supported important NASA missions, including gimbaled
optical systems for {\em WISE, HIRDLS, GIFTS,} and {\em GOES-R ABI,}
and SiC optical systems for {\em MICSAS, ALI, HIRDLS, LORRI,}
and {\em GOES-R ABI.}
L-3 IOS will build the CPI Science Payload drawing on its Tinsley
Laboratories facility for aspheric optics and L-3
Communications Systems East (L-3 CSE) for interface expertise
with the ISS. L-3 CSE was the primary contractor 
for the ISS communications and telemetry system with over \$600M in 
contracts. The CPI ICCD detector system will be procured from Mullard 
Space Science Laboratory, which cooperated in producing the CPI proposal.

The CPI Science Team includes Dan Reisenfeld who will act as Project 
Scientist and be responsible for scientific oversight of the CPI 
development, lead the laboratory and in-flight calibration and 
characterization effort including scientific performance tests, and 
participate in the data analysis. Paul Janzen will be the Instrument 
Scientist and will be responsible for carrying out the optical design 
and performance evaluation as well as the laboratory and in-flight 
calibrations and characterizations.
Steven Cranmer will act as Mission Scientist for Solar Wind
Science and John Raymond will be the Mission Scientist for CME Science.
The CPI Core Science Analysis Team will be greatly strengthened
by Ben Chandran, Terry Forbes, Alexander Panasyuk, Phil Isenberg,
and Aad van Ballegooijen. In addition to data analysis, 
Alexander Panasyuk will develop the ground software and head the 
science operations team.

\vspace*{-0.01in}
\begin{center}
{\em 6.2.  Technical Readiness Level and Risks}
\end{center}

\vspace*{-0.02in}
CPI will be built with existing 
technology that has extensive heritage. It has no medium or high risks. 
The Technical Readiness Levels (TRLs) of all CPI hardware is
at level 9 or higher. 

\vspace*{0.01in}
\begin{center}
{\em 6.3.  Technical Reserves, Cost Reserves, and Margin}
\end{center}

\vspace*{-0.02in}
Due to the substantial resource capabilities of ISS, 
mass, power, volume, and telemetry reserves are robust with 54\% 
reserves $+$ margin for mass, 376\% for power, 1000\% for uplink and 
downlink and a factor of 2 for pointing.

The CPI budget includes seven months of costed 
schedule reserve in addition to a schedule that provides 48 months from 
the start of phase B to the nominal launch date.
On top of all PI-Mission Costs and encumbered reserves,
there are unencumbered reserves of 25\% 
for phases A/B/C/D leaving 5\% margin to the Mission Cost Cap.
CPI has a robust descope plan with descopes across the project that 
would save approximately \$4M. The descopes increase risk by removing 
redundancy and by using a lower TRL command and data handling system, 
but the science impact is minimal. Phase E would be shortened by 10\% 
and loss in performance of optical reflection and detector efficiency 
would be accepted.
CPI has no foreign collaborations, contributions, or costs.
The ICCD detector system arrangement is a procurement, not an
international collaboration.

\vspace*{0.11in}
\begin{center}
{\bf REFERENCES}
\end{center}

\vspace*{-0.02in}
\parindent=0.0in
\baselineskip=12.93pt

Allen, L. A., et al. 2000, JGR, 105, 23123

Antiochos, S. K., et al. 2011, ApJ, 731, 112

Chandran, B. D. G. 2010, ApJ, 720, 548

Cranmer, S. R. 2009, Living Rev.\  Sol.\  Phys., 6, 3

Cranmer, S. R., et al. 1999, ApJ, 511, 481

Cranmer, S. R., et al. 2007, ApJS, 171, 520

Cranmer, S. R., et al. 2008, ApJ, 678, 1480

Cranmer, S. R., et al. 2010, Solar Phys., 263, 275

Cranmer, S. R., \& van Ballegooijen, A. A. 2005, ApJS, 156, 265

DeForest, C. E., \& Gurman, J. B. 1998, ApJ, 501, L217

De Pontieu, B., et al. 2007, Science, 318, 1574

Esser, R., \& Edgar, R. J. 2000, ApJ, 532, L71

Feldman, U., et al. 2007, ApJ, 660, 1674

Fisk, L. A. 2003, JGR, 108, 1157

Frazin, R. A., et al. 2003, ApJ, 597, 1145

Frazin, R. A., et al. 2009, ApJ, 701, 547

Hollweg, J. V., \& Isenberg, P. A. 2002, JGR, 107 (A7), 1147

Ko, Y.-K., et al. 2006, ApJ, 646, 1275

Ko, Y.-K., et al. 2010, ApJ, 722, 625

Kohl, J. L., et al. 1978, in New Instrumentation for Space
Astronomy (Oxford: Pergamon), 91

Kohl, J. L., et al. 1994, Space Sci.\  Rev., 70, 253

Kohl, J. L., et al. 1995, Solar Phys., 162, 313

Kohl, J. L., et al. 1996, ApJ, 465, L141

Kohl, J. L., et al. 1997, Solar Phys., 175, 613

Kohl, J. L., et al. 1998, ApJ, 501, L127

Kohl, J. L., et al. 1999, ApJ, 510, L59

Kohl, J. L., et al. 2006, A\&A Review, 13, 31

Kowal, G., et al. 2009, ApJ, 700, 63

Krishna Prasad, S., et al. 2011, A\&A, 528, L4

Kumar, A., \& Rust, D. M. 1996, JGR, 101, 15667

Laming, J. M. 2009, ApJ, 695, 954

Landi, E. 2008, ApJ, 685, 1270

Landi, E., et al. 2010, ApJ, 711, 75

Lee, J.-Y., et al. 2009, ApJ, 692, 1271

Lee, L. C., \& Wu, B. H. 2000, ApJ, 535, 1014

Lenz, D. D. 2004, ApJ, 604, 433

Lin, J., \& Forbes, T. G. 2000, JGR, 105, 2375

Lin, J., et al. 2004, ApJ, 602, 422

Lin, R. P., et al. 2010, {\em Solar Eruptive Events (SEE)
2020 Mission Concept,} white paper submitted to
NRC Solar/Space Physics Decadal Survey

Markovskii, S. A., et al. 2006, ApJ, 639, 1177

Mason, K. O., et al. 2001, A\&A, 365, L36

Moore, R. L., et al. 2011, ApJ, 731, L18

Murphy, N. A., et al. 2011, ApJ, in press, arXiv:1104.2298

Ofman, L., \& Davila, J. M. 2001, ApJ, 553, 935

Panasyuk, A. V. 1999, JGR, 104, 9721

Pasachoff, J. M., et al. 2007, ApJ, 665, 824

Raymond, J. C. 2008, J.\  Astrophys.\  Astron., 29, 187

Reeves, K. K., et al. 2010, ApJ, 721, 1547

Roming, P. W. A., et al. 2000, Proc.\  SPIE, 4140, 76

Scudder, J. D. 1992, ApJ, 398, 299

Sheeley, N. R., Jr., et al. 1997, ApJ, 484, 472

Sturrock, P. A., \& Coppi, B. 1966, ApJ, 143, 3

van der Holst, B., et al. 2010, ApJ, 725, 1373

Vi\~{n}as, A. F., et al. 2000, ApJ, 528, 509

Voitenko, Y., \& Goossens, M. 2004, ApJ, 605, L149

Wilhelm, K. 2006, A\&A, 455, 697

Withbroe, G. L., et al. 1982, Space Sci.\  Rev., 33, 17

Yang, S., et al. 2011, ApJ, 732, L7

Yashiro, S., et al. 2004, JGR, 109, A07105

Zurbuchen, T. H. 2007, Ann.\  Rev.\  Astron.\  Astrophys., 45, 297

\end{document}